# The Classification of Cropping Patterns Based on Regional Climate Classification Using Decision Tree Approach

Tb Ai Munandar and Sumiati

*Department of Informatics Engineering,*
*Faculty of Information Technology, Universitas Serang Raya, Banten, Indonesia*

**Abstract:** Nowadays, agricultural field is experiencing problems related to climate change that result in the changing patterns in cropping season, especially for paddy and coarse grains, pulses roots and Tuber (CGPRT/Palawija) crops. The cropping patterns of rice and CGPRT crops highly depend on the availability of rainfall throughout the year. The changing and shifting of the rainy season result in the changing cropping seasons. It is important to find out the cropping patterns of paddy and CGPRT crops based on monthly rainfall pattern in every area. The Oldeman's method which is usually used in the classification of of cropping patterns of paddy and CGPRT crops is considered less able to determine the cropping patterns because it requires to see the rainfall data throughout the year. This research proposes an alternative solution to determine the cropping pattern of paddy and CGPRT crops based on the pattern of rainfall in the area using decision tree approach. There were three algorithms, namely, J48, RandomTree and REPTree, tested to determine the best algorithm used in the process of the classification of the cropping pattern in the area. The results showed that J48 algorithm has a higher classification accuracy than RandomTree and REPTree for 48%, 42.67% and 38.67%, respectively. Meanwhile, the results of data testing into the decision tree rule indicate that most of the areas in DKI Jakarta are suggested to apply the cropping pattern of 1 paddy cropping and 1 CGRPT cropping (1 PS + 1 PL). While in Banten, there are three cropping patterns that can be applied, they are; 1 paddy cropping and 1 CGPRT cropping (1 PS + 1 PL), 3 short-period paddy croppings or 2 paddy croppings and 1 CGPRT cropping (3 short-period PS or 2 PS + 1 PL) and 2 paddy croppings and 1 CGPRT cropping (2 PS + 1 PL).

## Introduction

Farming activities in Asia play an important role, especially for the agrarian countries. Agriculture becomes an important source of foreign exchange for a country through the export of agricultural products, especially for paddy and CGPRT (palawija) crops. Therefore, there are various efforts made to improve the excellence of agricultural products in order to achieve maximum results in accordance with the expected target. One of the efforts made to realize excellence in the agricultural field is by recognizing the suitable cropping patterns based on climate indicators in each area.

Cropping patterns, especially for paddy and CGPRT crops, are usually known in three ways: (1) by checking the calendars and planting seasons in a one-year cycle, (2) by relying on the old methods to recognize the cropping season for paddy and CGPRT crops based on rainy season and (3) by using the Oldeman approach to find out the pattern of yearly cropping season based on the rainfall indicators. These three approaches, on the other hand, are not effective for current use. The yearly cropping calendar is usually made based on the rainy season pattern occurs. However, as climate change globally, it is difficult to predict the current rainy season since it often shifts from one month to another. It requires to renew the cropping season calendar regularly, so that it complicates the farmers to determine the suitable cropping patterns in the areas to be cropped by paddy and CGPRT crops.

The old approach used to find out the cropping season pattern is also not always effective due to the inconsistent impacts of climate change. The most possible way is by analyzing the cropping season pattern using the Oldeman approach. However, this approach also depends on the rainfall conditions of a area before producing the suitable cropping pattern for a particular area. The Oldeman analysis is only able to recognize the cropping pattern for the current condition. In addition, the determination of the cropping pattern using the Oldeman approach requires to classify the annual rainfall data based on wet, moist and dry months, not based on the real monthly rainfall data. Whereas, the monthly rainfall data actually have certain patterns that can be used to determine the cropping pattern. Therefore, there is a need of an approach that is able to integrate the analysis of the three approaches above, so that it is not only able to find out the current cropping pattern, but also able to identify the future cropping pattern. This research does not specifically discuss the agricultural products based on certain cropping patterns. This research attempts to propose a new approach to determine the cropping pattern of a area based on classification technique approach using decision tree. Therefore, this research begins with the selection of decision tree algorithm to be used in the classification of the cropping pattern. The selection of the model is conducted by testing three algorithms into training data to establish the decision tree. The three algorithms are J48, RandomTree and REPTree. The results of the test of three algorithms are then compared to obtain the better decision tree algorithm. Furthermore, the testing data were tested into the selected decision tree algorithm to classify the cropping pattern of the area in accordance with the rainfall pattern within.

## Literature Review

There have been many research conducted on the determination of cropping pattern. Some research determined the cropping pattern based on the old way, for example, by analyzing the rainfall conditions occurred in a area, as conducted by Sujalu *et al.* (2014), evaluating the cropping patterns based on the yields of previous paddy cropping to determine the suitable cropping pattern in the following cropping season (Naing *et al.*, 2002), managing the cropping season pattern based on the intensity of the rainy and post-rainy seasons by looking at soil moisture and residual nutrients left over other crops in the previous season (Singh *et al.*, 2014). Analyzing the cropping patterns by utilizing the data as conducted by Gumma *et al.* (2014) to map the cropping season of paddy using the analysis of moderate-resolution imaging spectroradiometer (MODIS) data.

Some of the research above still used the old approach in determining the cropping patterns, except the research conducted by Gumma *et al.* (2014). There were not many research conducted by utilizing the patterns of rainfall data to determine the cropping patterns in accordance with the regional climate. Again, this research was conducted to develop a new approach in determining the cropping patterns based on the yearly rainfall patterns of each area. The decision tree approach was used in this research.

## The Classification of Regional Climate Using Oldeman

Oldeman is a regional climate classification method based on the yearly rainfall indicators. The regional climate classification is based on the information regarding the number of wet and dry months according to the intensity of yearly rainfall of a area. Oldeman classified a area into five major types and four sub-types of climate (Sudrajat, 2009; Sasminto *et al.*, 2014). The results of the classification conducted by Oldeman are also able to provide recommendations on the suitable cropping patterns, especially for paddy and CGPRT crops (Dewi, 2005). The regional climate classification using Oldeman approach is shown in Fig. 1.

The regional climate classification using Oldeman provides recommendations on the suitable cropping patterns of paddy and CGPRT crops according to the identified climatic type. Table 1 shows the cropping patterns of paddy and CGPRT crops based on Oldeman.

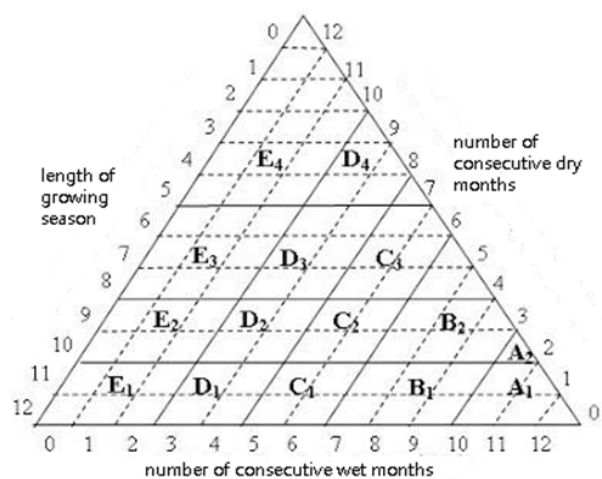

Fig. 1. Oldeman's triangle

Table 1. The Oldeman's Classification of Cropping Patterns of Paddy and CGPRT crops

| Climate type | Cropping pattern | Information |
|---|---|---|
| A1, A2 | Sustainably suitable for Paddy crops but has less production due to the low solar radiation flux density throughout the year | 3 short-period PS or 2 PS + 1 PL |
| B1 | Sustainably suitable for Paddy crops, with a good planning on the early cropping season. The production will be high in the dry-season harvest | 3 short-period PS or 2 PS + 1 PL |
| B2 | Capable for two paddy-cropping periods in a year with a short general variety and the short-period dry season which is able for CGPRT cropping | 2 PS + 1 PL |
| C1 | 1 period of Paddy cropping and 2 periods of CGPRT cropping in a year | 1 PS + 2 PL |
| C2,C3,C4 | Possible for 1 period of Paddy cropping and 2 periods of CGPRT cropping in a year. However, the second period of CGPRT cropping should be done carefully to avoid the dry season. | 1 PS + 1 PL |
| D1 | 1 short-period paddy cropping and it usually has high production due to the high solar radiation flux density. The perfect time to have CGPRT cropping. | 1 PS + 1 PL |
| D2,D3,D4 | It is only possible to have 1 period of paddy cropping and 1 period of CGPRT cropping in a year, depends on the availability of irrigating water. | 1 PS or 1 PL |
| E | The area with this type of climate is usually dry, which is only possible to have 1 period of CGPRT cropping, depends on the rainfall. | 1 PL |

## Decision Tree Model

Decision tree is the approach used for prediction analysis. It works based on the probability conditions of the gain and entropy values at the time of the establishment of the decision tree (Geetha and Nasira, 2014). Here are some decision tree methods which are often used for the establishment of decision tree, which are also used in this study.

### J48 Algorithm

J48 is a decision tree-establishing algorithm developed based on the ID3 concept (Gayatri *et al.*, 2009). The gain and gain ration values are usually used to determine the main root establishing the decision tree according to the existing attributes. The followings are the J48 algorithm:

- Choosing which attribute to be the root of the decision tree
- Creating the branches for each value
- Dividing the cases into the branches created in stage 2
- Repeat the process in every branch until all cases on the branches have the same classes

### RandomTree Algorithm

RandomTree algorithm is an approach used for the establishment of decision tree by considering a number of randomly-selected K attributes at each node formed. This algorithm does not cut the decision tree branches created as in other decision tree methods. It even provides the possibility of estimating the probability of a class based on its hold-out set. In this method, any attribute that is not selected to be a decision tree branch still has the opportunity to be a branch at the next level (Dhurandhar and Dobra, 2008).

### NBTree Algorithm

NBTree is a decision tree establishing algorithm which is generated from the provided dataset. The information regarding the frequency of class occurrence becomes the main basis for the establishment of decision tree. The determination of leave for each established rule is conducted by using the Naive Bayes technique. The NBTree algorithm is shown according to the following stages (Mori and Umezawa, 2009):

- Determining the early condition
- Classifying the data, then calculating each of the split nodes
- Evaluating the optimal tree and cross-validation error by cutting the established tree.
- Testing the established decision tree using the testing data
- Identifying each terminal node based on the testing data
- Predicting one step further using the Naïve Bayes on the resulted terminalnodes

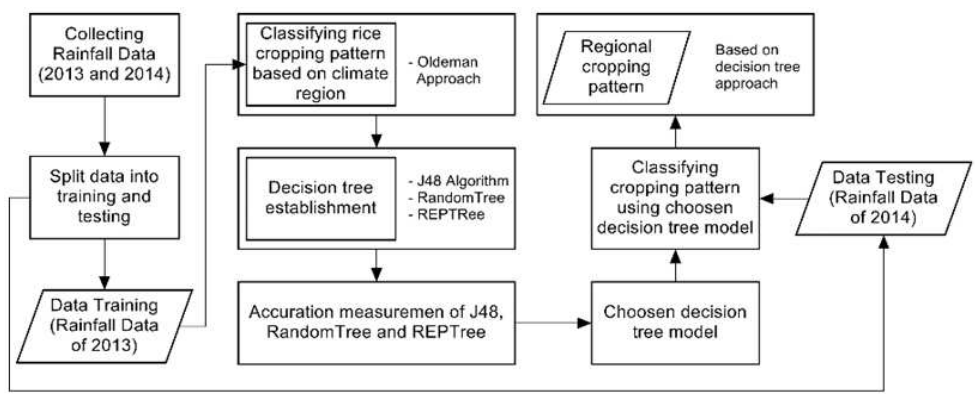

Fig. 2. Research stages

## Research Methodology

The research was started by collecting rainfall data from the Regional Climatology Agency of DKI and Banten Provinces for the data of 2013 and 2014. There were 75 rainfall data collection stations in total used as the data in this research. In this research, the rainfall data of 2013 was used as training data to establish the decision tree, while the rainfall data of 2014 was used as testing data. Weka software was used as the tool in this research. Before getting inputted into the decision tree model, there was a classification of the regional climate made from each station using the Oldeman approach. The Oldeman classification results were then used as training data and modeled into J48, Random Tree and REPTree to establish a decision tree. The decision tree models made of these three methods were then compared to see the best level of accuracy of the decision tree rules established. The testing data was then tested using the selected decision tree. The final results are in the form of cropping patterns for paddy and CGPRT crops based on the classification established from the decision tree. The research diagram is shown in Fig. 2.

## Results and Discussion

The initial stage of this research was by conducting climate classification using Oldeman. Climate classification was conducted towards the rainfall data 2013, collected from 75 rainfall collection stations. The results of the classification from 75 stations were divided into six types of regional climate. The six climate types are A1, A2, B2, B3, C3 and C4. Table 2 is according to the results of the regional climate classification using Oldeman. The findings of Oldeman were then incorporated into the decision tree models of J48, RandomTree and REPTree.

*The Comparison between the Decision Tree Models*

Decision trees were formed using three algorithms namely J48, RandomTree and REPTree.

Table 2. The Regional Climate Classification of The Rainfal Data of 2013

| Regional Climate type | Regional agency | |
|---|---|---|
| | DKI Jakarta | Banten |
| A1 | 0 | 8 |
| A2 | 3 | 16 |
| B1 | 0 | 0 |
| B2 | 6 | 14 |
| B3 | 2 | 15 |
| C1 | 0 | 0 |
| C2 | 0 | 0 |
| C3 | 0 | 9 |
| C4 | 0 | 2 |
| D1 | 0 | 0 |
| D2 | 0 | 0 |
| D3 | 0 | 0 |
| D4 | 0 | 0 |
| E | 0 | 0 |

Table 3. Comparison between J48, RandomTree and REPTree

| Indicators | J48 | RandomTree | REPTree |
|---|---|---|---|
| Classicifation Accuracy (%) | 48% | 42.67% | 38.67% |
| Kappa | 0.3363 | 0.2596 | 0.2109 |
| Mean absolute error | 0.1734 | 0.1911 | 0.2216 |
| Root mean square error | 0.3828 | 0.4372 | 0.3738 |
| Number of tree | 31.0000 | 53.0000 | 15.0000 |

The results of the decision trees established from each algorithm were then compared to determine the best algorithm used. The comparison of algorithms were based on several indicators, among others, the percentage of accuracy of decision tree establishment (classification accuracy), kappa value, absolute error mean value, root mean square error value and number of trees. Table 3 shows the comparison between J48, RandomTree and REPTree algorithms.

Based on the results of testing on the training data into the J48, RandomTree and REPTree algorithms, it was found that, in general, J48 algorithm has a better accuracy than the other two algorithms. It can be seen from the four comparative indicators as shown in Table 3.

Table 4. The differences in regional climate classification for 25 stations with incomplete data

| No | Regional Agency | Oldeman | J48 | Random Tree | REP Tree |
|---|---|---|---|---|---|
| 1 | Kedoya | B2 | C3 | C4 | C3 |
| 2 | Manggarai | B2 | C3 | C4 | C3 |
| 3 | Sunter III Rawabadak | A2 | C3 | C4 | C3 |
| 4 | Sunter Kodamar | B2 | C3 | C4 | C3 |
| 5 | BPP Caringin | A2 | C3 | C4 | C3 |
| 6 | UPTD Kronjo | C3 | C3 | C4 | C3 |
| 7 | UPTD Rajeg Banyawakan | B3 | C3 | C3 | C3 |
| 8 | UPTD Sindang Jaya | C3 | C3 | C4 | C3 |
| 9 | Baros | B2 | B2 | C3 | C3 |
| 10 | Singamerta | C3 | C3 | C4 | C3 |
| 11 | Tirtayasa | C4 | C3 | C4 | C3 |
| 12 | Walantaka | C3 | C3 | C4 | C3 |
| 13 | Bojong | B3 | C3 | C4 | C3 |
| 14 | Cikeusik | B3 | C3 | C4 | C3 |
| 15 | Cimanggu | C3 | C3 | C4 | C3 |
| 16 | Jiput | A1 | C3 | C4 | C3 |
| 17 | Mandalawangi | B3 | C3 | C4 | C3 |
| 18 | Pagelaran | B2 | C3 | C4 | C3 |
| 19 | Bojong Manik | A2 | C3 | C4 | C3 |
| 20 | BPP Leuwidamar | B2 | C3 | C4 | C3 |
| 21 | Cijaku | A1 | C3 | C4 | C3 |
| 22 | Cijaura/ Cimesir | A2 | C3 | C4 | C3 |
| 23 | Cisangu Atas | B3 | C3 | C4 | C3 |
| 24 | Sampang Peunduey | A2 | C3 | C4 | C3 |
| 25 | Warung Gunung | B3 | C3 | C4 | C3 |

Table 5. The recommended cropping patterns for DKI Jakarta areas

| No | Regional Agency | Climate Class | Cropping Patterns |
|---|---|---|---|
| 1 | Halim | B2 | 2 PS + 1 PL |
| 2 | Kemayoran | C3 | 1 PS + 1 PL |
| 3 | Tanjung Priok | C3 | 1 PS + 1 PL |
| 4 | Karet | B2 | 2 PS + 1 PL |
| 5 | Kedoya | C3 | 1 PS + 1 PL |
| 6 | Manggarai | C3 | 1 PS + 1 PL |
| 7 | Pakubuwono | A2 | 3 PS short-period or 2PS + 1 PL |
| 8 | Pulogadung | C3 | 1 PS + 1 PL |
| 9 | Rorotan | C3 | 1 PS + 1 PL |
| 10 | Sunter III Rawabadak | C3 | 1 PS + 1 PL |
| 11 | Sunter Kodamar | C3 | 1 PS + 1 PL |

The comparative indicators of classification accuracy show that the percentage of the accuracy of J48 algorithm classification is higher than RandomTree and REPTree. Similarly, for the Kappa value, J48 has a higher value than the other two algorithms. In the calculation of the error value, the root mean square error value of J48 algorithm is not better than REPTree but better than RandomTree. Meanwhile, for the mean absolute error value, J48 algorithm is still better than RandomTree and REPTree.

The visualization of the decision trees established from the three algorithms has some significant differences. In the J48 algorithm, for example, from twelve month attributes, there were only nine attributes used as the criteria to establish the branches of a tree. Meanwhile, for the RandomTree algorithm, there were only eleven month attributes used as the criteria to establish the branches of the decision tree. The significant difference was found in the REPTree algorithm, from twelve month attributes, there were only five month attributes used as the criteria to establish the branches of the tree.

The next comparison was conducted the two results of testing on the testing data. The test on the testing data was conducted in two stages. The first stage was conducted on the rainfall collected from 75 stations without considering the stations which do not have complete rainfall data or do not have any rainfall data at all.

The second stage was conducted after reducing the station data which do not complete rainfall data or do not have any data at all. In the second stage of testing, there were 51 stations tested with complete rainfall data. The results of the first stage test show that the difference of accuracy level for each algorithm is not very significant.

Table 6. The recommended cropping pattern for Banten areas

| No | Regional Agency | Climate Class | Cropping Patterns | No | Regional Agency | Climate Class | Cropping Patterns |
|---|---|---|---|---|---|---|---|
| 1 | Cengkareng | C3 | 1 PS + 1 PL | 33 | Pontang | C3 | 1 PS + 1 PL |
| 2 | Curug | A2 | 3 PS short-period or 2PS + 1 PL | 34 | Ragas Hilir | C3 | 1 PS + 1 PL |
| 3 | Pondok Betung | B2 | 2 PS + 1 PL | 35 | Singamerta | C3 | 1 PS + 1 PL |
| 4 | Serang | C3 | 1 PS + 1 PL | 36 | Tirtayasa | C3 | 1 PS + 1 PL |
| 5 | Tangerang | C3 | 1 PS + 1 PL | 37 | Walantaka | C3 | 1 PS + 1 PL |
| 6 | BPP Caringin | C3 | 1 PS + 1 PL | 38 | Bd Ciliman | A2 | 3 PS short-period or 2PS + 1 PL |
| 7 | Jatiwaringin Mauk | C3 | 1 PS + 1 PL | 39 | Bojong | C3 | 1 PS + 1 PL |
| 8 | UPTD Balaraja | B2 | 2 PS + 1 PL | 40 | Cibaliung | A2 | 3 PS umur pendek atau 2 PS + 1 PL |
| 9 | UPTD Benda Sukamulya | B2 | 2 PS + 1 PL | 41 | Cikeusik | C3 | 1 PS + 1 PL |
| 10 | UPTD Bendung Ciputat | B2 | 2 PS + 1 PL | 42 | Cimanggu | C3 | 1 PS + 1 PL |
| 11 | UPTD Cipondoh | A2 | 3 PS short-period or 2PS + 1 PL | 43 | Cimanuk | A1 | 3 PS short-period or 2PS + 1 PL |
| 12 | UPTD Kresek | A2 | 3 PS short-period or 2PS + 1 PL | 44 | Jiput | C3 | 1 PS + 1 PL |
| 13 | UPTD Kronjo | C3 | 1 PS + 1 PL | 45 | Labuhan | A2 | 3 PS short-period or 2PS + 1 PL |
| 14 | UPTD Rajeg Banyawakan | C3 | 1 PS + 1 PL | 46 | Mandalawangi | C3 | 1 PS + 1 PL |
| 15 | UPTD Sepatan | C3 | 1 PS + 1 PL | 47 | Menes | A2 | 3 PS umur pendek atau 2 PS + 1 PL |
| 16 | UPTD Serpong | A2 | 3 PS short-period or 2PS + 1 PL | 48 | Pagelaran | C3 | 1 PS + 1 PL |
| 17 | UPTD Sindang Jaya | C3 | 1 PS + 1 PL | 49 | Pandeglang | A2 | 3 PS short-period or 2PS + 1 PL |
| 18 | UPTD Tegal Kemiri | C3 | 1 PS + 1 PL | 50 | Bojong Leles | A2 | 3 PS short-period or 2PS + 1 PL |
| 19 | Anyer | C3 | 1 PS + 1 PL | 51 | Bojong Manik | C3 | 1 PS + 1 PL |
| 20 | Baros | B2 | 2 PS + 1 PL | 52 | BPP Leuwidamar | C3 | 1 PS + 1 PL |
| 21 | Carenang | C3 | 1 PS + 1 PL | 53 | BPP Sajira | A2 | 3 PS short-period or 2PS + 1 PL |
| 22 | Cinangka | B2 | 2 PS + 1 PL | 54 | Cijaku | C3 | 1 PS + 1 PL |
| 23 | Ciomas | A2 | 3 PS short-period or 2PS + 1 PL | 55 | Cijaura/ Cimesir | C3 | 1 PS + 1 PL |
| 24 | Ciruas | C3 | 1 PS + 1 PL | 56 | Cilaki/ Ciminyak | B2 | 2 PS + 1 PL |
| 25 | Kasemen Kilasah | C3 | 1 PS + 1 PL | 57 | Cisangu Atas | C3 | 1 PS + 1 PL |
| 26 | Kragilan Kalenpetung | B2 | 2 PS + 1 PL | 58 | Kecamatan Cimarga | A2 | 3 PS short-period or 2PS + 1 PL |
| 27 | Kramatwatu Pegadingan | C3 | 1 PS + 1 PL | 59 | Lebak Parahiang | A2 | 3 PS short-period or 2PS + 1 PL |
| 28 | Mancak | B2 | 2 PS + 1 PL | 60 | Malingping Utara | A2 | 3 PS short-period or 2PS + 1 PL |
| 29 | Pabuaran | A2 | 3 PS short-period or 2PS + 1 PL | 61 | Panyaungan | A2 | 3 PS short-period or 2PS + 1 PL |
| 30 | Padarincang | B2 | 2 PS + 1 PL | 62 | Pasir Ona Rangkas | A2 | 3 PS short-period or 2PS + 1 PL |
| 31 | Pamarayan | A2 | 3 PS short-period or 2PS + 1 PL | 63 | Sampang Peundeuy | C3 | 1 PS + 1 PL |
| 32 | Petir | B2 | 2 PS + 1 PL | 64 | Warung Gunung | C3 | 1 PS + 1 PL |

The accuracy levels for J48, RandomTree and REPTree algorithms are 17.33%, 12.00% and 14.67%, respectively. The second stage of testing has an increase in the level of accuracy of each algorithm. However, the J48 algorithm still has a better accuracy than other algorithms. The accuracy levels in the second stage of testing for the J48, RandomTree and REPTree algorithms are 25.49%, 17.65% and 21.57%, respectively. For the stations with empty rainfall data, the result of the test in the first stage shows that J48 and REPTree algorithms identified it as the C3-type climate, while the RandomTree result of the station with empty rainfall data identified it as the C4-type climate. The differences of the regional climate classification for the stations with incomplete data are shown in Table 4.

*Regional Cropping Pattern Classification*

The cropping pattern classification was conducted by inputting testing data into selected algorithm with the highest accuracy. In this research, based on the test results, the J48 algorithm was selected for its decision tree to determine the cropping pattern of a area. The decision tree was then put in a breakdown to generate a rule in determining the regional climate pattern based on monthly rainfall numbers of each area.

The result of the regional climate classification was then analyzed based on the cropping pattern based on Table 1. For the area of DKI Jakarta, the suitable cropping pattern for its surrounding stations is dominated by the pattern of 1 period of paddy cropping and 1 period of CGPRT cropping (1 PS + 1 PL), similarly for the Kemayoran, Tanjung Priok, Kedoya, Manggarai, Pulogadung, Rorotan, Sunter III Rawabadak and Sunter Kodamar areas. While for areas such as Halim and Karet, it is suggested to have a cropping pattern with a combination of 2 periods of paddy cropping and 1 period of CGPRT cropping (2 PS + 1 PL). Meanwhile, for Pakubuwono area, the recommended cropping pattern is 3 periods of short-period paddy cropping or 2 periods of paddy cropping and 1 period of CGPRT cropping. The cropping patterns for DKI Jakarta and surrounding areas are shown in Table 5.

The cropping patterns for Banten areas showed that most of the stations and surrounding stations are suggested to use the cropping pattern of 1 period of paddy cropping and 1 period of CGPRT cropping (1 PS + 1 PL). This applies to several stations and surrounding areas, such as, Cengkareng, Serang, Tangerang, Anyer, Ciruas and Kasemen. The second cropping pattern for some other areas is suggested to have 3 periods of short-period paddy cropping or 2 periods of paddy cropping and 1 period of CGPRT cropping (3 short PS or 2 PS + 1 PL). Some areas suggested to use this second cropping pattern are Ciomas, Pabuaran, Pamarayan, Cibaliung, Labuhan, Menes and Cimanuk. The last cropping pattern is 2 periods of paddy cropping and 1 period of CGPRT cropping (2 PS + 1 PL). The areas which are more suitable to use this cropping pattern include Pondok Betung, Balaraja, Baros, Cinangka, Kragilan, Mancak and Padarincang. Table 6 shows the complete cropping pattern for every area in Banten.

## Conclusion

Based on the research, it can be concluded that J48 algorithm has better accuracy than RandomTree and REPTree. The results of the test on the testing data into J48 showed that most of the regional stations and surrounding areas in DKI Jakarta are suggested to apply the cropping pattern of 1 period of paddy cropping and 1 period of CGPRT cropping. For Banten, some stations have several different cropping patterns. More than a half of the areas in Banten is suggested to have the cropping pattern of 1 period of paddy cropping and 1 period of CGPRT cropping. One third of Banten area is suggested to have a cropping pattern of 3 periods of short-period paddy cropping or 2 periods of paddy cropping and 1 period of CGPRT cropping. While the one sixth remaining is suggested to have a cropping pattern of 2 periods of paddy cropping and 1 period of CGPRT cropping. The results of cropping pattern classification using decision tree can be used as an alternative approach to determine the cropping pattern, especially for paddy and CGPRT/palawija crops, in the future. With the decision tree J48 the classification of cropping pattern is conducted by looking at the pattern of monthly rainfall directly. Unlike the Oldeman which has to classify the monthly rainfall data first before classifying the regional climate.

This research can be used as an alternative recommendation for government in agriculture and even agriculture industry. Unpredictable climate change can be predicted early in particular to determine the timing of rice and crops. Thus the possibility of harvest failure can be minimized.

## Acknowledgement

Acknowledgements are conveyed to the Directorate General of Research and Development The Ministry of Research, Technology and Higher Education who has funded this research.

## Author's Contributions

**Tb Ai Munandar:** Participated in data analysis using Oldeman, decision tree and test the accuracy of grouping results using decision tree techniques.

**Sumiati:** Participated in collecting rainfall data collected from 65 rainfall collection stations. In addition, he contributed to check the accuracy of the results of grouping results using both Oldeman and decision trees.

## Ethics

No ethical issues would arise after the publication of this manuscript.